\documentclass[conference,twoside,letterpaper]{IEEEtran}

\ifCLASSINFOpdf
  \usepackage[pdftex]{graphicx}
  \graphicspath{{./figures/}}  
  \DeclareGraphicsExtensions{.pdf,.jpeg,.png}
\else
  \usepackage[dvips]{graphicx}
  \graphicspath{{./figures/}}  
  \DeclareGraphicsExtensions{.eps}
\fi

\usepackage{amsmath}
\usepackage{amssymb}

\usepackage{cite}

\usepackage{url}

\usepackage[dvipsnames]{xcolor}
\usepackage{bm}
\usepackage{cespvect}

\usepackage{flushend}

\newcommand{\figw}{1.0\columnwidth}
\newcommand{\figwmedium}{0.7\columnwidth}

\newcommand{\be}{\begin{equation}}
\newcommand{\ee}{\end{equation}}
\newcommand{\bea}{\begin{eqnarray}}
\newcommand{\eea}{\end{eqnarray}}
\newcommand{\bc}{\begin{cases}}
\newcommand{\ec}{\end{cases}}
\newcommand{\bi}{\begin{itemize}}
\newcommand{\ei}{\end{itemize}}

\newcommand{\ts}{\mathbf{x}}
\newcommand{\cts}{\hat{\ts}}
\newcommand{\model}{\mathbf{y}}
\newcommand{\fv}{\mathbf{f}}
\newcommand{\nfv}{\hat{\fv}}
\newcommand{\dm}{\mathbf{F}}
\newcommand{\ndm}{\hat{\dm}}
\newcommand{\median}{\mathrm{median}}
\newcommand{\iqr}{\mathrm{iqr}}

\newcommand{\mypar}[1]{\noindent \textbf{#1}}

\newtheorem{mydef}{Definition}

\begin{document}
%

\title{\mbox{Rate-Distortion} Classification for \\ Self-Tuning IoT Networks}


\author{Davide~Zordan, Michele~Rossi, Michele~Zorzi\\
Dept. of Inf. Engineering, via Gradenigo 6/b, 35131, Padova, Italy 
\vspace{-0.5cm}
}

\maketitle


\begin{abstract}
Many future wireless sensor networks and the Internet of Things are expected to follow a software defined paradigm, where protocol parameters and behaviors will be dynamically tuned as a function of the signal statistics. New protocols will be then injected as a software as certain events occur. For instance, new data compressors could be (re)programmed \mbox{on-the-fly} as the monitored signal type or its statistical properties change. We consider a lossy compression scenario, where the application tolerates some distortion of the gathered signal in return for improved energy efficiency. To reap the full benefits of this paradigm, we discuss an automatic sensor profiling approach where the signal class, and in particular the corresponding \mbox{rate-distortion} curve, is automatically assessed using machine learning tools (namely, support vector machines and neural networks). We show that this curve can be reliably estimated \mbox{on-the-fly} through the computation of a small number (from ten to twenty) of statistical features on time windows of a few hundreds samples. 
\end{abstract}


%
\IEEEpeerreviewmaketitle

\thispagestyle{empty}

\section{Introduction}
\label{sec:intro}

Wireless Sensor Networks (WSNs) and their evolution into the so called Internet of Things (IoT) have been extensively investigated in the last few years. Many papers have appeared on data gathering~\cite{Akkaya-2005} and signal processing~\cite{Qina-2016}. A challenging problem to face in distributed and large sensing systems is to efficiently disseminate the gathered data, by jointly performing some sort of signal processing to reduce as much as possible the size of the packets that are being transmitted, while retaining most of the information therein or, better stated, preserving the information that is required by the application, no less, no more. This efficient data collection translates into energy savings for the sensor nodes, that are often battery operated and thus require a careful design of every task. This is usually achieved through the application of signal compression, often lossy~\cite{Zordan-2014}, in both space and time~\cite{Quer-2012,Hooshmand-2016} and also through packet aggregation at the relay nodes, as the packets travel through the network~\cite{Fasolo-2007}. So, increasing the energy efficiency of distributed sensor networks often entails the use of joint signal processing (e.g., compression) and dissemination techniques (e.g., transmission and data aggregation), which must be properly tuned according to the statistics of the \mbox{(spatio-temporal)} signal that is being measured. In this paper, we discuss a novel paradigm for the automatic tuning of these algorithms, based on {\it sensor profiling} through data mining. 

Data mining is becoming increasingly important~\cite{Tsai-2014} and here we advocate a new methodology that exploits these techniques to automatically adjust WSN/IoT signal compression and data gathering protocols. Our point is that in the near future most of the network functionalities and the network behavior itself will be {\it injected as a software} into the network, according to a software defined paradigm. \mbox{Off-the-shelf} sensor hardware will be used for most \mbox{(non-critical)} applications, while its actual configuration will be carried out after deployment. As a clarifying example, consider a door open/close sensor. This hardware detects the door's activity and both the sensing hardware and the type of signal it measures are very simple. Imagine  that we are confronted with the problem of assessing with which technique, and how much, the signal generated by this sensor has to be compressed, i.e., picking the right compression algorithm and selecting the most appropriate \mbox{rate-compression} tradeoff. One may think that this problem is promptly solved by simply hardcoding the type of signal being measured into the sensor hardware and then using a compression algorithm that is known to be good for that signal. While this in general may provide an acceptable solution to our problem, we believe that it is possible to do much better. In fact, the signal statistics (i.e., number of open/close events) per unit time and, in turn, the type of traffic generated by the sensing hardware depends on {\it where} the door sensor was installed and {\it when} the signal is measured. Clearly, a door sensor in the front door of a public building will generate a rather intense activity in the daytime, but the signal from a closet door will probably show a small number of events. During the night, both sensors are likely to show no events at all. Similar considerations hold for parking sensors, i.e., different locations have different parking behaviors depending on the type of street (e.g., residential {\it vs} shopping neighborhoods), and time of the day.

So, our point is that knowing {\it what} is being measured is often not enough, but we would ideally want to acquire more knowledge about the signal statistics such as the temporal and spatial correlation of the data or, even more, its \mbox{\it rate-distortion} characteristics. Having that, we could automatically control how much the signal can be compressed, by still meeting the application requirements in terms of reconstruction quality (fidelity). Compression can then be performed at the network edge (right at the field sensors)~\cite{Zordan-2014} or through a distributed approach (\mbox{in-network} processing)~\cite{Fasolo-2007}.

In this paper, we present a first step toward the automatic classification of sensor signals, showing an application example to assess its potential in terms of energy savings for the IoT network. Our aim is to reliably predict the \mbox{rate-distortion} function of a generic temporal signal by analyzing a small window of samples. We foresee a usage model where data is gathered and, upon collecting a few samples (e.g., $500$ samples are used in our results), the time series can be automatically classified in terms of \mbox{rate-distortion} behavior for selected compression algorithms. Having this function, or at least a good estimate of it, makes it possible to decide upon the most suitable compression algorithm to use and to automatically tune it. Besides compression at the source, the estimated \mbox{rate-distortion} tradeoff can be exploited to design and/or adapt channel access and routing protocols, which shall be jointly optimized with the compression algorithm. In this paper we first focus on the problem of reliably assessing \mbox{rate-distortion} curves, exploring computationally intensive as well as lightweight approaches, and evaluating their performance with a large number of time series from diverse domains. Hence, we assess the energy savings that arise from the application of this paradigm within a \mbox{single-hop} IEEE~802.15.4 network. 

Our chief goal is to show that a \mbox{data-driven} sensor profiling for the automatic configuration of network parameters and protocols is in fact a feasible and practical approach, which can also be executed at a relatively low computational cost exploiting certain classifiers, such as neural networks.  

The rest of this paper is organized as follows. In Section~\ref{sec:signals} we discuss the considered signal types. In Section~\ref{sec:signal_compression} we focus on the compression algorithms, providing definitions for the rate and distortion metrics. Data mining techniques for sensor profiling are presented in Section~\ref{sec:signal_classification}, their performance is evaluated in Section~\ref{sec:results} and a suitable application scenario is investigated in Section~\ref{sec:application_scenario}. Our concluding remarks and future research directions are discussed in Section~\ref{sec:conclusions}.

\section{Signals}
\label{sec:signals}

For the purpose of this paper, we collected diverse univariate real world time series, which were acquired from publicly available datasets and from our own measurement campaign of vital signs (mainly electrocardiogram, respiratory and heart rates). These have been selected as a representative set of the signals types that are acquired in common IoT scenarios, including: 1) environmental sensing~\cite{Sensorscope}, e.g., temperature, humidity, soil moisture, precipitation measures, wind speed, solar radiation, 2) biomedical applications~\cite{Saeed-2011,Arrhythmia,Physionet}, e.g, electrocardiograms (ECG), photoplethysmograms, respiration signals, 3) smart electricity grids and smart cities, e.g., power consumption of home appliances \cite{Smart} and measures of building's structural strain. In total, we have run experiments on $7,010$ time series taken from these application domains. 

Every signal in the database is sampled at a constant rate; this rate is \mbox{signal-specific} but our aim is to come up with algorithms that are agnostic to it. For our experiments, we divided each time series into \mbox{non-overlapping} temporal windows of $N$ samples, so that each time window is an array of $N$ real values, i.e., \mbox{$\ts = (x_1 , x_2 , \dots , x_N ) \in \mathbb{R}^N$}. In the remainder of this paper, with ``input time series'' we refer to one such window of data for a specific signal type. This is convenient because the considered compression and data mining algorithms operate on time sequences of $N$ samples.     

\begin{figure}[tb]
	\centering
	\includegraphics[width=0.8\columnwidth]{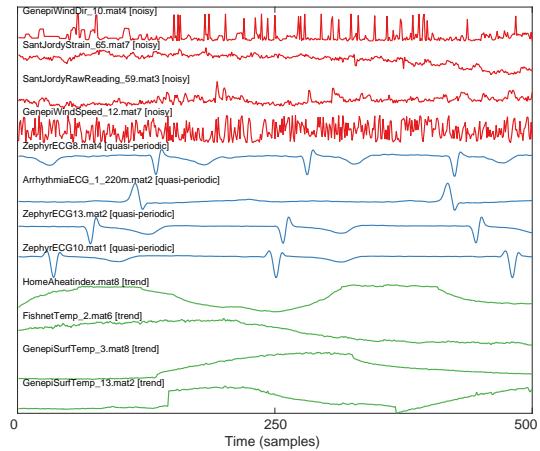}
	\caption{Four randomly selected time series for each of the three signal classes.}
	\label{fig:class_example}
\end{figure}

Based on preliminary results on compression schemes, and considering the analysis in~\cite{Zordan-2014}, we group the signals into three classes, namely: {\it i)} {\bf noisy signals}, such as wind speed and structural strain, where the temporal correlation is low and the time series show an erratic behavior, which is difficult to predict and has no evident trend or periodic components; {\it ii)} {\bf quasi-periodic signals}, such as ECG traces and other biomedical time series, where a similar pattern is repeated over time, with small variations in  shape and duration; and {\it iii)} {\bf trend signals}, such as measures of temperature, humidity, and other environmental quantities, that exhibit a slowly varying behavior and have a noticeable trend component. In Fig.~\ref{fig:class_example}, we qualitatively show some example time series (windows of $N=500$ samples each) for each signal class (four examples per class): the noisy signals are on top (red color in the figure), \mbox{quasi-periodic} signals in the middle (blue) and trend ones at the bottom (green).    

Signals in each class are supposed to perform similarly when going through the process of (temporal) compression. Moreover, each time series, after being classified, can be associated with a certain \mbox{rate-distortion} curve, which is representative of the class it belongs to. Thus, this curve can be used to optimize the operation of networking protocols, e.g., to minimize the energy expenditure entailed by the data collection algorithms, given an error tolerance for the signal reconstructed at the WSN data collector (the {\it sink}).

The method that we discuss in this paper to perform this classification task uses a combination of feature extraction and machine learning techniques, and is detailed in Section~\ref{sec:signal_classification}. 

\section{Signal Compression} 
\label{sec:signal_compression}

From the analysis in~\cite{Zordan-2014}, we consider two lossy compression algorithms, which are suitable for IoT \mbox{sensing-and-report} applications, namely, {\it i)} Lightweight Temporal Compression (LTC)~\cite{Schoellhammer-2004} and {\it ii)} compression based on Discrete Cosine Transform (DCT). LTC is among the most lightweight compression techniques for WSNs, whereas DCT-based algorithms usually provide the best accuracy, but are more energy demanding. Both compression schemes take as input the data to compress, $\ts$, and an error parameter, $\varepsilon$, and output a model $\model$ for the compressed signal. The model is then transmitted and used at the sink to obtain the reconstructed signal $\cts$. A brief description of the LTC and DCT compression algorithms is provided next.\\

\begin{figure*}[tb]
	\centering
	\includegraphics[width=1.8\columnwidth]{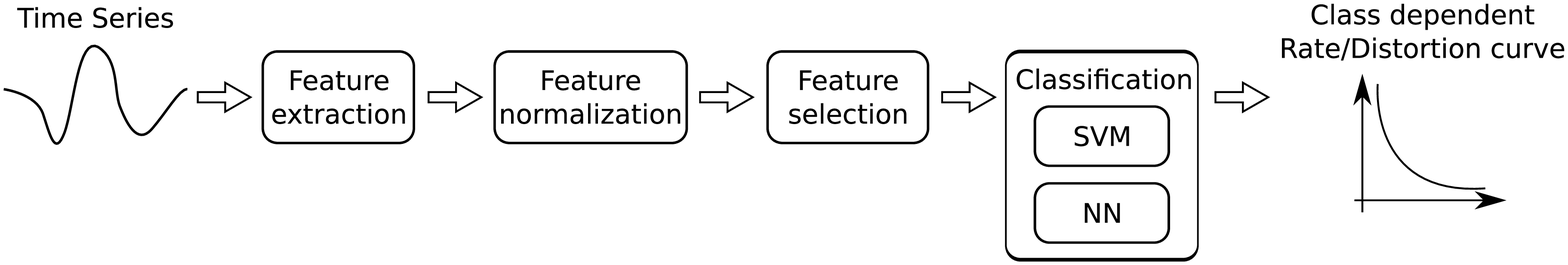}
	\caption{Flow diagram of the proposed data mining framework.}
	\label{fig:block_scheme}
\end{figure*}

\mypar{LTC} approximates the time series $\ts$ by means of linear segments. The first segment is built starting from the first two samples of $\ts$, then subsequent samples are iteratively added and the segment slope is updated. This is iterated as long as the segment approximates all the samples from the beginning of it with an error smaller that $\varepsilon$. Once it is no longer possible to meet the error constraint using the current segment, a new segment is initiated and the two end points of the previous one are saved in the model $\model$. The procedure is repeated until the last sample of the time series is reached.\\

\mypar{DCT} approximates the time series $\ts$ by retaining a fraction of the coefficients of its DCT transform $\mathbf{X}$. DCT is selected over other frequency domain transforms because of its good properties, in particular: {\it i)} its coefficients are real; {\it ii)} it has a strong ``energy compaction'' property, i.e., the signal's energy tends to be concentrated in a few low-frequency components and {\it iii)} it does not suffer from the edge discontinuity problem, which is instead typical of other techniques. The compression algorithm works as follows: the time series $\ts$ is moved into the frequency domain through DCT, the model $\model$ is built by retaining the first coefficient of $\mathbf{X}$ and zero padding to the length of $\ts$ (coefficients are added from the lowest to the highest frequency). Hence, at the compressor, the reconstructed signal $\cts$ is obtained through the inverse DCT of $\model$ and the error constraint is checked for each sample of $\cts$. If the error constraint is met the algorithm stops, otherwise, the next frequency coefficient is added to $\model$ and the procedure is iterated.\\

The two compression algorithms are applied to the input time series by varying the error parameter $\varepsilon$ and returning an empirical \mbox{rate-distortion} curve for each of the input signals and for each compression algorithm. These curves are then grouped by signal class and the average \mbox{rate-distortion} curve for each class is evaluated (example \mbox{rate-distortion} functions are discussed shortly through Figs.~\ref{fig:rd_ltc} and~\ref{fig:rd_dct}).
The formal definitions of the concepts of rate and distortion that we use in this work are provided next.\\
\begin{mydef}[Rate]
	Given a time series $\ts$ and its compressed representation $\model$, we define the compression \textbf{rate} as:
	\begin{equation}
		\eta = \frac{N_b(\model)}{N_b(\ts)} \; , \; \eta \in (0,1] \; ,
		\label{eq:eta}
	\end{equation}
	where $N_b(\ts)$ and $N_b(\model)$ are the number of bits required to represent the original time series $\ts$ and the compressed signal model $\model$, respectively.\\
\end{mydef}

\begin{mydef}[Distortion]
	Given a time series $\ts$ and its reconstructed version $\cts$, we define the \textbf{distortion} as:
	\begin{equation}
		\hat{\varepsilon} =  \frac{\max_{i=1,\dots,N}\left\lbrace\left| x_i - \hat{x}_i\right|\right\rbrace}{\max_i\{ x_i\} - \min_i \{ x_i\}}\cdot 100  \; ,
	\end{equation}
	which corresponds to the maximum distance between the samples of $\ts$ and $\cts$, normalized to the range of the values in the original time series $\ts$.
\end{mydef}

\section{Signal Classification} 
\label{sec:signal_classification}
The data mining procedure that we have developed is based on the extraction of features from the original time series. In the following, we describe the feature set, how it is computed and how the feature-based representation of a time-series can be used to classify the corresponding signal. In particular, in Section~\ref{sec:feature_extraction} we discuss the {\it feature extraction} procedure, in Section~\ref{sec:feature_normalization} we describe the {\it feature normalization} step, and in Section~\ref{sec:feature_selection} we present a method to reduce the number of features to be extracted (i.e., the {\it feature selection} block) and subsequently used for {\it classification}, which we discuss in Section~\ref{sub:classification}. A flow diagram of the proposed data mining approach is provided in Fig.~\ref{fig:block_scheme}.


\subsection{Feature Extraction} 
\label{sec:feature_extraction}

The feature extraction phase is performed through the Highly Comparative Time Series Analysis (HCTSA) framework~\cite{Fulcher-2014}. This framework includes a large collection of methods for time-series analysis and allows converting a time series into a vector of (thousands of) informative features, each obtained from a specific operation on the temporal signal. Diverse operations are accounted for and include basic statistics of the distribution of time series values (e.g., location, spread, Gaussianity, outlier properties), linear correlations (e.g., autocorrelations, power spectrum analysis), stationarity (e.g., StatAv, sliding window measures, prediction errors), information theoretic and entropy/complexity measures (e.g., auto-mutual information, Approximate Entropy, Lempel-Ziv complexity), methods from nonlinear time series analysis (e.g., correlation dimension, Lyapunov exponent estimates, surrogate data analysis), linear and nonlinear model fits (e.g., goodness of fit estimates and parameter values from autoregressive moving average, ARMA, Gaussian Process, and generalized autoregressive conditional heteroskedasticity, GARCH, models), and many others (e.g., wavelet methods, properties of networks derived from time series).

Each operation in the framework is encoded as an algorithm taking as input a time series $\ts = (x_1, x_2, \dots , x_N)$, and returning a single real number $f_i$, that is denoted as a feature. The collection of all the output features for an input time series is referred to as {\it feature vector} $\fv = (f_1, f_2, \dots , f_M) \in \mathbb{R}^M$. 

\subsection{Feature Normalization} 
\label{sec:feature_normalization}

Using features for the classification of time series requires the definition of a proper distance metric between any two feature vectors. Moreover, when such a large number of operations with different output distributions is involved, a transformation that allows a meaningful comparison of feature vectors is also required.
In particular, when calculating distances between feature vectors, the output range of all the operations in the framework should be similar, so that all operations are equally weighed. Although many different normalization approaches may be used, one that is simple and robust to the presence of outliers in the distribution of operation outputs is the {\it \mbox{outlier-robust} sigmoidal transform}: 
\begin{equation}
	\nfv = \left\lbrace 1 + \exp \left[-\frac{\fv - \median (\fv)}{1.35 \cdot \iqr (\fv)}\right] \right\rbrace^{-1} \; ,
	\label{eq:normalization}
\end{equation}
where the use of the median and the inter quartile range of $\fv$ ($ \iqr (\fv)$ in the equation), instead of the mean and variance, makes the transformation less sensitive to outliers and other artifacts in the distribution of $\fv$. Upon applying the nonlinear transformation in Eq.~(\ref{eq:normalization}), the results are linearly scaled to the interval $[0,1]$ so that every operation (feature) has the same output range.

\subsection{Feature Selection} 
\label{sec:feature_selection}

In many supervised classification applications the set of features that is extracted from the original data can be reduced to a smaller subset of highly relevant features. This process is called {\it feature selection} and is useful to reduce the computational burden of the feature extraction task, as well as to ameliorate and to possibly eliminate overfitting. Although many different feature selection methods have been proposed in the literature, such as the lasso~\cite{Lasso-1994}, elastic net~\cite{Zou-2005}, and recursive feature elimination~\cite{Guyon-2002}, HCTSA implements a transparent and easily interpretable feature selection method called {\it greedy forward feature selection}. According to this technique, we iteratively select the features that maximize the classification accuracy of a linear classifier, adding one feature at a time to the final feature subset. Despite being suboptimal with respect to considering multiple features jointly, this greedy approach is lightweight and was found to provide satisfactory results. 

\subsection{Classification} 
\label{sub:classification}

We consider a supervised classification approach. There, each input time series has an entangled feature vector and a desired output value, i.e., the label corresponding to the class the time series belongs to.

In this paper, we consider two different classifiers, namely: 1) a linear Support Vector Machine (SVM) classifier, and 2) a Feed Forward Neural Network (FFNN) classifier. 
Both these methods are trained using a stratified $k$-fold cross validation approach~\cite{Kohavi-1995}. Specifically, the original sample data is divided into $k$  subsamples of the same size, $k-1$ of the subsamples are used as training data, and the remaining subsample is used as validation data to test the classification accuracy. The $k$ subsamples are randomly selected from the sample data, retaining the same proportion of instances per signal class in each subsample. The cross validation process is repeated $k$ times, with each subsample used exactly once as test data, and the classification accuracy is obtained averaging the results of the $k$ cross validation iterations. 

The multiclass linear SVM classifier is built using \mbox{$C(C-1)/2$} binary \mbox{one-versus-one} linear SVMs, where $C$ is the number of classes.
For the neural network classifier, we used a FFNN with a single hidden layer and sigmoid neurons~\cite{Bishop-2007}. 

\section{Classification and Rate-Distortion Results} 
\label{sec:results}

For our experiments we considered a set of over $7,000$ time series of fixed length $N = 500$ samples, obtained from consecutive non overlapping portions of the databases described in Section~\ref{sec:signals}. We have first run the feature extraction procedure on the input time series applying all the operations in the HCTSA library, which outputs about $9,000$ features per time series. After this processing phase, the time series and the operations that produced errors or special valued outputs (e.g., NaN, Inf, etc.) have been filtered out, leaving us with a set of $S = 6,707$ time series and $M = 5,254$ features. These were stored in a {\it signal-feature matrix}, $\dm \in \mathbb{R}^{S\times M}$, where each row contains the feature vector associated with a specific time series in the dataset. Each column of $\dm$ represents a different feature across the whole dataset, that is then normalized using Eq.~(\ref{eq:normalization}), and the normalized features are finally stored in the \mbox{signal-feature} matrix $\ndm$.

\begin{figure}[tb]
	\centering
	\includegraphics[width=\figw]{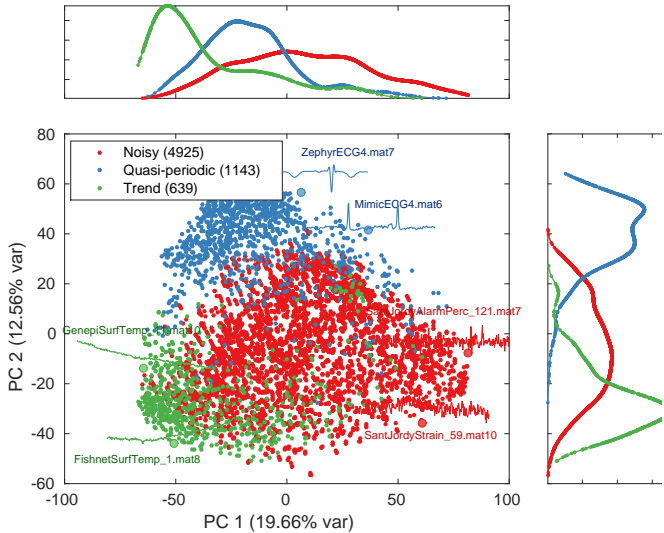}
	\caption{First two principal components of the signal feature vectors for each signal class and the corresponding empirical pdfs (top and right plots).}
	\label{fig:pca_plot}
\end{figure}

A basic means of visualizing a low-dimensional representation of $\ndm$ is shown in Fig.~\ref{fig:pca_plot}, where we apply Principal Component Analysis (PCA) on the feature set for each signal class. Specifically, we consider all the signals in a certain class, along with the corresponding feature vectors, and use PCA to obtain the two leading Principal Components (PC) of $\ndm$. Each point in Fig.~\ref{fig:pca_plot} represents one of the $S$ input time series, and the color is associated with the class the time series belongs to. The axis labels correspond to the first two Principal Components, termed PC1 and PC2, and the proportion of variance in the dataset that is described by each PC is indicated between parentheses. Moreover, the two small graphs on the top and to the right-hand side of Fig.~\ref{fig:pca_plot} show the marginal distribution (probability density function, pdf) of the two PCs, for each signal class. Remarkably, the three classes are already quite distinguishable by just using two PCs. In fact, points of the same color are mostly grouped together, even though there is some overlapping in the central part of the pdf plots. This is however only meaningful as a visual inspection tool, but is still not enough to automatically classify signals.   

\begin{figure}[tb]
	\centering
	\includegraphics[width=\figw]{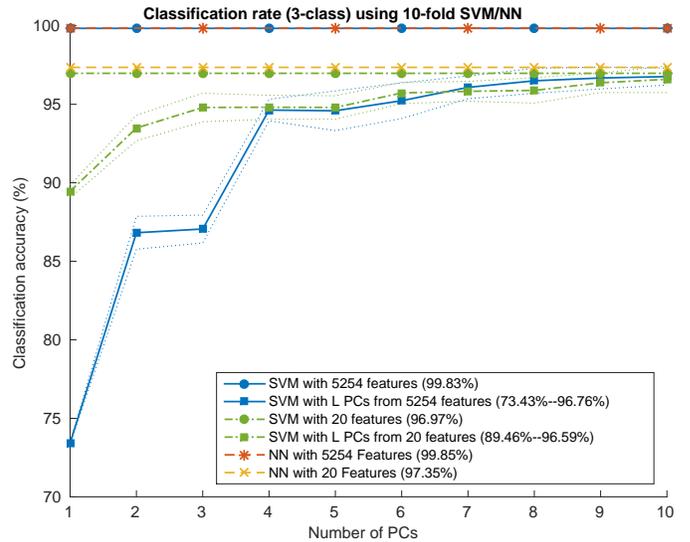}
	\caption{Classification accuracy using a $10$-fold cross validation approach for: 1) an SVM and an FFNN classifier trained on all the $5,254$ features extracted (i.e., on the entire matrix $\ndm$), 2) an SVM classifier trained on the $L$ principal components of $\ndm$, 3) SVM and FFNN classifiers trained on the twenty most representative features, selected through a greedy procedure and 4) an SVM classifier trained on $L \leq 10$ principal components of the reduced and normalized $S\times 20$ \mbox{signal-feature} matrix.}
	\label{fig:accuracy}
\end{figure}

As discussed in Section~\ref{sec:signal_classification}, automatic classifiers are instead obtained by training SVMs and FFNNs using the features in $\ndm$, and evaluating the classification accuracy for each class. This is done by labelling each signal as belonging to a certain class and then using this label for the training, through a supervised approach. Fig.~\ref{fig:accuracy} shows that using either an SVM or an FFNN classifier, when we consider all the $5,254$ features the classification accuracy is very high, i.e., higher than $99.8 \%$. The use of the full feature set is however computationally demanding and impractical, especially if this classification task has to be carried out at the network edge (i.e., at the sensor nodes). We thus have to substantially reduce the number of relevant features to compute, in the hope that this will still lead to high classification rates. 

Toward this end, in Fig.~\ref{fig:accuracy} we also show the classification accuracy we obtain when we train an SVM classifier using the first $L$ PCs of $\ndm$, with $1 \le L \le 10$, which grows from $73.43\%$ using just the first PC to $96.76 \%$ when using the first ten PCs. As an example, using the information contained in Fig.~\ref{fig:pca_plot} (i.e., only two PCs) to train an SVM classifier, we can get a classification accuracy of roughly $87 \%$. Computing feature vectors of ten elements (the first ten PCs) is certainly appealing, but this still entails the fact that the whole feature set has to be obtained first, which is computationally expensive.  

For this reason, we applied the greedy feature selection scheme of Section~\ref{sec:feature_selection} to select the twenty most representative features from the $5,254$ that were originally extracted; this returns a reduced and normalized $S \times 20$ \mbox{signal-feature} matrix. As noted above, this feature selection procedure is heuristic and better feature sets may be extracted through more involved (but computationally demanding) approaches, nonetheless it allows to extract only twenty features from the original dataset, considerably reducing the processing cost. We then trained SVM and FFNN classifiers using the so identified twenty features and the corresponding classification results are also shown in Fig.~\ref{fig:accuracy}. As we can see from this plot, SVM and FFNN classifiers in this case lead to similar accuracies of about $97 \%$. Note that this accuracy is even higher than that achieved by considering all the $5,254$ features and using the first ten PCs computed on the whole feature set to train the classifiers. The reason behind this is that with too many features we risk to overfit the data and, in turn, our classifiers will have a worse generalization capability than those built on a much smaller but highly representative feature set.\footnote{This is a common and well known problem in data analysis, see, e.g., the discussion in~\cite[Chapter 1]{Bishop-2007}.} As a last curve, in this plot we show the classification accuracy obtained by training the classifiers on the first $L \leq 10$ PCs of the reduced $S \times 20$ \mbox{signal-feature} matrix. As expected, the classification accuracy increases with an increasing number of PCs, getting very close to that obtained with the best twenty features when the first ten PCs are considered. 

\begin{figure}[tb]
	\centering
	\includegraphics[width=\figwmedium]{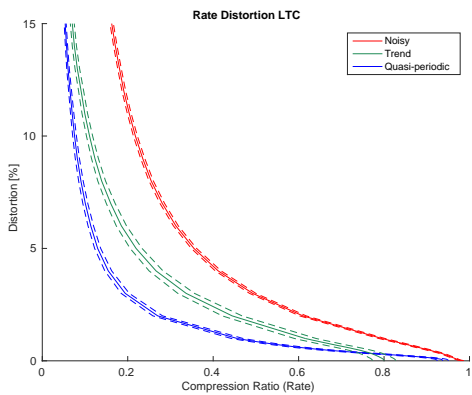}
	\caption{Rate-Distortion curves for the considered signal classes, Lightweight Temporal Compression (LTC) algorithm.}
	\label{fig:rd_ltc}
\end{figure}

\begin{figure}[tb]
	\centering
	\includegraphics[width=\figwmedium]{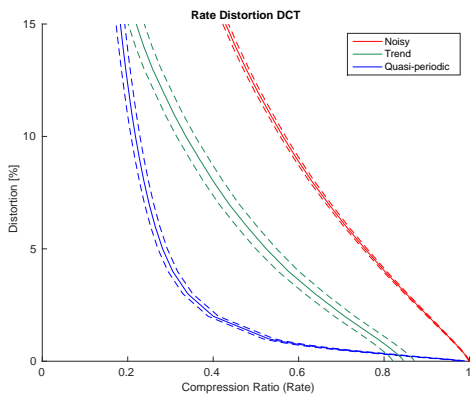}
	\caption{Rate-Distortion curves for the considered signal classes, Discrete Cosine Transform (DCT) compression algorithm.}
	\label{fig:rd_dct}
\end{figure}

As a final result, Figs.~\ref{fig:rd_ltc} and~\ref{fig:rd_dct} show the average rate distortion curves corresponding to the three signal classes, for LTC and DCT, respectively. These curves were obtained by averaging the \mbox{rate-distortion} points within each class. It is evident that there are substantial differences in the compression performance for signals that belong to different classes. Hence, knowing the class a signal belongs to, or being able to classify the signal with high accuracy, allows estimating the expected performance in terms of \mbox{rate-distortion} behavior, and this makes it possible to infer how much distortion we will get by reducing the data we transmit through compression. For example, with LTC a maximum distortion requirement of $4$\% entails a maximum compression of $\eta=0.5$ for noisy signals, as opposed to $\eta=0.2$ with \mbox{quasi-periodic} ones. From Eq.~(\ref{eq:eta}), $\eta=0.5$ means that the number of bits that the sensors transmit is \mbox{one-half} of those that were sampled, whereas with $\eta=0.2$ only \mbox{one-fifth} of the data has to be transmitted, which entails a considerable reduction in the transmission energy. We expect the application to dictate the maximum tolerable distortion and the \mbox{compression-transmission} protocols to adapt to it at runtime, as the signal statistics change.

\section{Application Scenario} 
\label{sec:application_scenario}

Next, we present an example application scenario to illustrate the benefits of automated IoT signal classification in terms of reduced energy consumption at the nodes, under a certain QoS constraint, i.e., error tolerance in the signal reconstruction at the data collector. We consider a star topology network featuring IEEE~802.15.4 compliant nodes which periodically sense data (sampling) and send reports to a data collector (the sink). We assume the following: the sampled signals belong to one of the three classes of Section~\ref{sec:signals}, the sink is not power constrained and can thus classify them through one of the methods of Section~\ref{sec:signal_classification}, the IEEE~802.15.4 beaconless mode with CSMA/CA Medium Access Control (MAC)~\cite{IEEE-802-15-4} is exploited for data transmission and the send times for the reports are randomized to reduce the MAC collision probability.

The application tolerates a maximum reconstruction error $\xi$, so we let the nodes apply DCT compression to reduce the amount of data to send to the sink: this entails a higher energy efficiency at the cost of an increased error in the reconstructed signal. As a  benchmark scenario, we first evaluate the energy consumption when the nodes send their data without compression. We then compare this case against two further configurations: 1) \textbf{DCT classless compression} (DCT-CL), where each node compresses its own data using a compression ratio from an average rate distortion curve ($\xi$ is plugged into that curve to obtain the corresponding minimum rate); 2) \textbf{DCT class aware compression} (DCT-CA), where each node is first classified\footnote{Note that the classification task can be performed by the sink after the first report period, where the nodes may send the data without compressing it. The class information can be sent back to the sensors via a control message or an acknowledgment packet.} and then uses a compression ratio obtained through the assigned {\it class specific} rate distortion curve. 

\begin{figure}[tb]
	\centering
	\includegraphics[width=0.85\columnwidth]{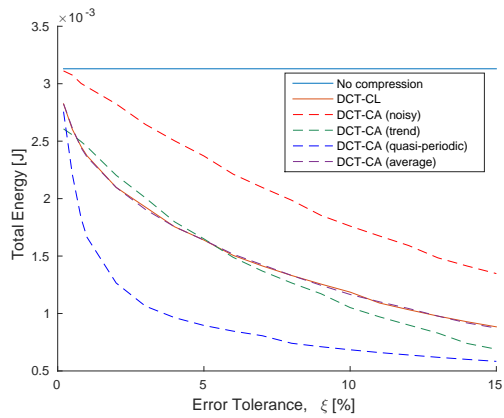}
	\caption{Energy consumption per report period \emph{vs} error tolerance dictated by the application.}
	\label{fig:energy}
\end{figure}

\begin{figure}[tb]
	\centering
	\includegraphics[width=0.85\columnwidth]{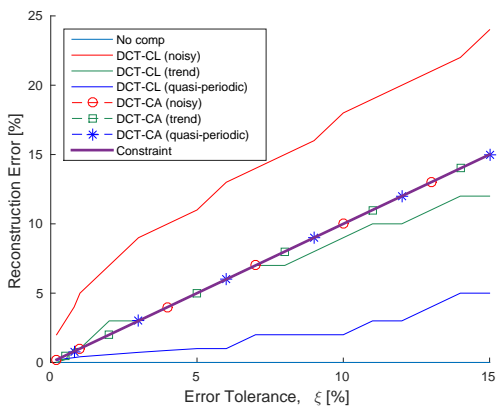}
	\caption{Reconstruction error per report period \emph{vs} error tolerance dictated by the application.}
	\label{fig:error}
\end{figure}

Fig.~\ref{fig:energy} and~\ref{fig:error} show the energy consumption per report period and the reconstruction error at the sink as a function of the error tolerance $\xi$ respectively. The energy consumption is evaluated adding up the energy spent for transmission and compression within a report period. In the former, we include the cost associated with the transmission of IEEE~802.15.4 packet headers (multiple packets are used when the amount of data to be transmitted in a round exceeds the maximum IEEE~802.15.4 payload size). We assume that each signal sample takes $16$ bits in fixed-point number representation, and we set the sampling interval to $10$ seconds. Nodes store $N=500$ samples before sending a report, so the report period is roughly $83$ minutes.

The maximum energy drainage is attained when the data is sent uncompressed (``No-compression'' in the plots). Here, the energy consumption is constant and the corresponding reconstruction error is zero. When applying DCT-CL the total energy decreases for an increasing application error tolerance, but the corresponding reconstruction error in some cases exceeds the application constraint $\xi$. This is because the compression ratio is picked using an average rate-distortion curve, which may provide rough estimates of the actual \mbox{rate-distortion} performance for a specific signal class. With DCT-CA, each node picks its compression ratio using the rate-distortion curve of the signal class the node belongs to. This yields energy savings for all the nodes but in different amounts. In fact, in order to meet the application constraint $\xi$, nodes can apply different compression levels depending on their specific class, and this entails higher transmission costs for nodes that can compress less and vice-versa. Nodes that are classified as {\it noisy} experience the lowest energy reduction, while {\it trend} and {\it quasi-periodic} ones achieve energy savings similar to those of DCT-CL (trend class) or higher (quasi-periodic). Also, the actual reconstruction error of DCT-CA always lies within the constraint $\xi$ dictated by the application.

\section{Conclusions and Future Research Directions}
\label{sec:conclusions}

In this paper, we have discussed a data mining framework to automatically assess the \mbox{rate-distortion} curve of WSN signals through lightweight classifiers. Our objective is to use these curves to adapt protocol parameters \mbox{on-the-fly} or to inject new protocol functionalities as the signal type or its statistics undergoes major changes. Our preliminary experiments, conducted on real signals, are encouraging and indicate that a small number of features (smaller than $20$) suffices to achieve a satisfactory classification performance (higher than $97$\%). An application example is finally provided to quantify the energy savings that are allowed by a correct classification, while also proving that a single ``average'' rate-distortion curve does not suffice.

A number of related research directions is currently under investigation. We would like to increase the classification granularity within each signal class. Also, we also would like to account for \mbox{event-based} signals and we wish to characterize the computation complexity of the classifiers to assess whether it is feasible to execute them on sensor hardware. Finally, we need to check the quality of our estimates in dynamic scenarios, where the signal statistics change as a function of time, and assess the energy savings for the transmission protocols in this case.

\section*{Acknowledgment}

This work was supported in part by Intel's Corporate Research Council (``An energy- and context-centric optimization framework for IoT nodes,'' EC-CENTRIC).

\bibliographystyle{IEEEtran}
\bibliography{paper}

\end{document}